\documentclass[11pt,letter,subeqn,fleqn]{article}
\pdfoutput=1

\usepackage{amsmath, amssymb,amsthm,cite}
\usepackage[normal]{subfigure}
\usepackage{subfig}
\usepackage{graphicx}
\usepackage{float}
\usepackage[utf8]{inputenc}
\usepackage{color}
\usepackage{url}

\title{A Symbolic Computation Framework for Constitutive Modelling Based On Entropy Principles}

\author{ 
A. F. Cheviakov \footnotemark[1],~~J. Heß \footnotemark[2]\vspace{0.5cm}\\
\small $^{\rm a}$\emph{Department of Mathematics and Statistics, University of Saskatchewan, Saskatoon, Canada}\vspace{0.2cm}\\
\small $^{\rm b}$\emph{Department of Fluid Dynamics, Technical University of Darmstadt, Darmstadt, Germany}\vspace{0.2cm}\\ }

\def\beq{\begin{equation}}
\def\eeq{\end{equation}}
\def\barr{\begin{array}{ll}}
\def\earr{\end{array}}

\def\const{\hbox{\rm const}}

\newcommand{\p}{\partial\,}

\providecommand{\keywords}[1]{\textbf{\textit{Keywords:}} #1}

{\theoremstyle{definition}

}

\newcounter{tabnum}

\def\V{{\mathcal V}}

\newtheorem{lemma}{Lemma}

\begin{document}

\renewcommand{\baselinestretch}{1.0}\small\normalsize

\footnotetext[1]{Alternative English spelling: Alexey Shevyakov. Electronic mail: shevyakov@math.usask.ca}
\footnotetext[2]{Corresponding author. Electronic mail: hess@fdy.tu-darmstadt.de}


\maketitle
\numberwithin{equation}{section}

\begin{abstract}

The entropy principle in the formulation of M\"{u}ller and Liu is a common tool used in constitutive modelling for the development of restrictions on the unknown constitutive functions describing material properties of various physical continua. 

In the current work, a symbolic software implementation of the Liu algorithm, based on \verb|Maple| software and the \verb|GeM| package, is presented. The computational framework is used to algorithmically perform technically demanding symbolic computations related to the entropy principle, to simplify and reduce Liu's identities, and ultimately to derive explicit formulas describing classes of constitutive functions that do not violate the entropy principle. Detailed physical examples are presented and discussed.
\end{abstract}

\medskip \keywords{Constitutive Modelling, Entropy Principle, Symbolic Computations}

\section{Introduction}

Entropy principles are used in continuum mechanics in order to investigate the material behavior. For a given model, the universal balance laws, such as those for mass, momentum and energy, are commonly given by a system of partial differential equations (PDEs). The specific material behavior is defined by a \textit{constitutive model}, through the specification of \textit{constitutive functions} present in the system. While the set of balance equations holds for a wide class of physical settings, for example, gases, ideal and non-ideal fluids, elastic and plastic solids, etc., the constitutive functions prescribe individual material behavior. From the mathematical point of view, they provide closure conditions for the system of balance equations, so that its fields can be uniquely determined. For an overview of constitutive modelling in the context of continuum mechanics, see, e.g., Hutter and Jöhnk's extensive work \cite[p. 139ff]{hutter_continuum_2004}. 

The principles of constitutive modelling may vary depending on the application; they can be based on theoretical considerations, experimental data, and/or heuristic assumptions. Fundamental theoretical principles for the formulation of material models include the requirements of \textit{material objectivity}, \textit{material symmetry} and \textit{thermodynamic consistency}. The first requirement determines that material behavior, and therefore the constitutive equations, must be independent of the observer. The second rule points towards the fact that the material laws must also satisfy the symmetric properties of a body, such as invariance under rotations, translations, or Galilei transformations.

In this work, following the ideas of M\"{u}ller \cite{muller_thermodynamic_1968} and Liu \cite{liu_method_1972}, we focus on the third requirement, which demands that the constitutive functions are restricted in such a way that an \emph{entropy principle} holds for all solutions of the model. The entropy principle is formulated in terms of an entropy inequality expressing the second law of thermodynamics (SL). The requirement that the entropy production inequality holds for every solution yields a set of constraints on the model's constitutive functions. M\"{u}ller's approach is based on the fact that the generalized entropy inequality is linear with respect to a set of independent higher derivatives of field variables; the constraints arise as coefficients of such derivatives set to zero.

Multiple alternative approaches to constitutive modelling exist. For example, constitutive modelling approaches also can be based on the descriptions of material behavior by balance equations. For this, an additional class of fields is considered, called the \textit{internal variables}, accounting for an internal state of the material. Such variables represent the microstructure and physical mechanisms within the body, and their evolution is described by respective balance equations. An example is given by the volume fraction (e.g., \cite{goodman1972continuum}), which accounts for the microstructure and the distribution of a granular material. In such approaches, the entropy principle of M\"{u}ller and Liu provides an additional useful insight into the possible interdependence between those newly introduced internal variables and the general constitutive functions of a system.\\
Other approaches in modelling micromechanics apply a distinction between different scales, in which the microscale models are employed to generate macroscale constitutive functions through the method of homogenization. This was applied, for example, in \cite{goda2012micropolar,goda20143d} for anistropic micropolar continua, i.e., structured solids, in the context of continuum mechanics, but without a reference to the entropy principle. It should be noted that, while those approaches are more similar to those of so-called averaging theories in the context of mixtures of fluids, we stay within the limits of mixture theory, as explained below.

It should be noted that, as there exists a wide range of definitions of entropy, there are also multiple entropy principles. In the context of continuum mechanics, especially the approach of Coleman and Noll \cite{coleman_thermodynamics_1963} is worth mentioning, giving an entropy principle on the basis of the Clausius-Duhem inequality. Wang and Hutter \cite{wang_comparison_1999} pointed out, however, that for mixtures, for structured continua and for polar continua like solids or liquid crystals, the entropy principle of M\"{u}ller and Liu is to be favored. Furthermore, to the present day, the entropy principle of M\"{u}ller and Liu is applied in many different fields of modelling be it chemical processes \cite{reis2016two} or granular flows \cite{hess2017thermodynamically}.

Liu \cite{liu_method_1972} (also see \cite{liu2002shih}) systematized M\"{u}ller's procedure, applying the method of Lagrange multipliers. Liu's algorithm is significantly more general, and can be applied to a wide range of models, without the requirement that external supply terms be related. It can also be used for models that do not involve the physical entropy. In Section \ref{sec:EntropyPrinciple:ML}, we review the details of the problem of constitutive modelling based on entropy principles, and the main steps of M\"{u}ller's approach and the Liu algorithm, illustrating them for a specific example of an anisotropic heat-conducting fluid.

The Liu algorithm \cite{liu_method_1972} is based on the following lemma, formulated for linear algebraic equations and a linear inequality (see also \cite{liu2002shih} and \cite{hauser2002historical}).
\begin{lemma}[Liu] \label{lem:Liu}
Let $z\in \mathbb{R}^p$, and let $M$ be a $p\times n$ real matrix. Consider a linear system   $MY +z = 0$ of $p$ equations on the components of the unknown vector $Y\in \mathbb{R}^n$, with a non-empty solution set $S$.  Let also $\mu\in \mathbb{R}^n$, $\mu\ne 0$, and $\zeta\in \mathbb{R}$ be given. Then the following statements are equivalent:
\begin{enumerate}
  \item $\forall Y\in S$, $\mu^T Y + \zeta \geq 0$;
  \item $\exists \lambda \in \mathbb{R}^p$ such that $\forall Y \in \mathbb{R}^n$, $\mu^T Y + \zeta  - \lambda^T(MY +z) \geq 0$;
  \item $\exists \lambda \in \mathbb{R}^p$ such that $\mu=M^T\lambda$, and $\zeta  \geq \lambda^T z$.
\end{enumerate}
\end{lemma}
As pointed out in \cite{hauser2002historical}, Lemma \ref{lem:Liu} is a related to the fundamental inequality lemma of Farkas and Minkowski. The latter plays a key role in linear programming, and is in turn related to the general Hahn-Banach separation theorem.

The Lemma \ref{lem:Liu} has a different flavor from the Lagrange multiplier approach to constrained optimization of nonlinear functions. In particular, both the equations and the inequality in the problem are linear. Moreover, the geometrical meaning of the Lemma can be understood as follows: since the inequality $F=\mu^T Y + \zeta \geq 0$ must hold for \emph{all} points $Y$ in the set defined by $MY +z = 0$, that set (a line, a hyperplane, etc.) must be, in a certain sense, parallel to isosurfaces $F=\const.$

%

In the following Section \ref{sec:EntropyPrinciple:ML}, we review the main steps of the original M\"{u}ller method and the Liu algorithm involving Lagrange multipliers to review their suggested principles of constitutive modelling. A model of a simple heat conducting compressible anisotropic fluid (Section \ref{sec:RunEg}) is used as a running example to outline the stages.

The technical computations related to the execution of the M\"{u}ller-Liu procedure can be time consuming, and equations that arise tend to be quite lengthy. While simplest examples can be carried out within minutes by an experienced researcher, generally, the derivation of constraints on the constitutive functions can become error-prone, especially for complicated settings, such as mixture models involving multiple phases and internal variables, governed by additional balance equations, and/or having complex material behavior. One of the goals of the current paper is the application of modern  symbolic software to facilitate computations related to the Liu algorithm, in particular, lengthy chain rule differentiations, computation of coefficients at higher-order derivatives, and efficient reduction and solution of overdetermined systems of partial differential equations for the unknown constitutive functions. The computations are based on a symbolic package \verb|GeM| for \verb|Maple|, developed in \cite{cheviakov2007gem,cheviakov2010computation,cheviakov2010symbolic} for symmetry and conservation law computations. Significant similarities between the nature of those problems and the algorithm of Liu, and the capabilities of \verb|GeM| software to efficiently handle linear and nonlinear PDEs, partial derivatives of field variables, and constitutive functions that may involve derivatives, make \verb|Maple| and \verb|GeM| a natural computational platform choice.

One of the goals of the current paper is the application of modern powerful symbolic software to facilitate computations related to the Liu algorithm, in particular, lengthy chain rule differentiations, computation of coefficients at higher-order derivatives, and efficient reduction and solution of overdetermined systems of partial differential equations for the unknown constitutive functions. The computations are based on a symbolic package \verb|GeM| for \verb|Maple|, developed in \cite{cheviakov2007gem,cheviakov2010computation,cheviakov2010symbolic} for symmetry and conservation law computations. Significant similarities between the nature of those problems and the algorithm of Liu, and the capabilities of \verb|GeM| software to efficiently handle linear and nonlinear PDEs, partial derivatives of field variables, and constitutive functions that may involve derivatives, make \verb|Maple| and \verb|GeM| a natural computational platform choice.


\medskip The rest of the paper is organized as follows. In Section \ref{sec:EntropyPrinciple:ML}, we review the main steps of the original M\"{u}ller method and the Liu algorithm involving Lagrange multipliers to review their suggested principles of constitutive modelling. A running example of a model of a simple heat conducting compressible anisotropic fluid is used to illustrate the stages.


Section \ref{sec:Maple} outlines the main stages of the symbolic computation algorithm. Two examples of symbolic computations of entropy principle constraints are subsequently considered:  an elementary example of one-dimensional gas dynamics model (Section \ref{sec:Maple:EG1}), and a model of simple heat conducting compressible anisotropic fluid (Section \ref{sec:Maple:EG2}).

The paper is concluded with a discussion Section \ref{sec:discussion} containing remarks about theoretical and computational aspects of entropy principles and research directions.

%
\section{The Entropy Principle of M\"{u}ller and Liu for Constitutive Modelling}\label{sec:EntropyPrinciple:ML}

\subsection{The Problem of Constitutive Modelling}\label{sec:ConstitModProb}

Consider the motion of a physical continuum within the domain $\Omega\in \mathbb{R}^n$, $n\geq 1$. Specific instances include but are not limited to elastic media, ideal and non-ideal fluids, gases, and plasmas, as well as various kinds of mixtures, including those that undergo interactions and chemical reactions. The independent variables of the model are given by the time $t$ and the spatial variables (commonly Cartesian coordinates) $x = (x_1,\ldots, x_n)\in \mathbb{R}^n$ (see also Section \ref{sec:discussion}). Depending on the application, the spatial independent variables may be Eulerian or Lagrangian coordinates, the former choice being common in gas, fluid and plasma dynamics, and the latter choice in solid mechanics. In one, two and three spatial dimensions, the Cartesian spatial variables may be denoted, without confusion, by $x$, $(x,y)$, and $(x,y,z)$, respectively.

The physical parameters describing the continuum depend on time and spatial coordinates; they are given by the fields
\beq\label{eq:gen:PhysFields}
\phi=(\phi_1(t, x),\ldots,\phi_m(t, x)),
\eeq
which are the dependent variables of the model. The evolution of these quantities is described by the PDEs of the model,
\beq\label{eq:gen:PhysPDEs}
\Pi^{\phi} = \left\{\Pi^{\phi_1}, \Pi^{\phi_2},..,\Pi^{\phi_m}\right\}
\eeq
that include the balance equations and possibly additional constraints (such as, for example, the condition of a fluid being irrotational, or a plasma being field-aligned).

In addition to independent and dependent variables, mathematical models of continua commonly involve constitutive functions, which must be specified through constitutive equations. Such equations provide the mathematical closure of the governing equations, describing  the material behavior of a specific medium within the general class of materials whose dynamics obeys the balance equations $\Pi^{\phi}$.  As remarked in M\"{u}ller's paper \cite{muller_thermodynamic_1968}, constitutive equations may be nonlocal in time, for example, they may depend on the history of the process. Similarly, constitutive models may involve spatial integral or delay-type terms, etc. In the simplest, most commonly considered situations, the constitutive functions $\psi$ are given by local expressions
\beq\label{eq:def:constit}
\psi = \psi \left(\phi_C\right),
\eeq
depending on the set of variables $\phi_C$ that may include independent and dependent variables of the model, as well as possibly some specific the derivatives of the dependent variables.



\emph{The main problem of constitutive modelling} consists in a description of classes of the dependencies \eqref{eq:def:constit}, which affirm, or do not contradict, certain physical, mathematical, or philosophical principles, as well as possibly the available experimental data. The principles of constitutive modelling include:
\begin{itemize}
  \item Coordinate invariance, i.e., invariance with respect to coordinate transformations. This property is also referred to as the principle of \emph{material objectivity}; see, e.g., \cite[p. 148]{hutter_continuum_2004}.
  \item Material symmetry (e.g., \cite[p. 155]{hutter_continuum_2004}).
  \item Physical postulates and simplifying assumptions of physical and mathematical nature.
\end{itemize}

%
\subsection{A Running Example}\label{sec:RunEg}

Throughout the current section, as a running example, we will use the model of a simple heat conducting compressible anisotropic fluid, following M\"{u}ller \cite{muller_thermodynamic_1968} and Liu \cite{liu_method_1972}. The main physical flow parameters are the density $\rho=\rho(x,t)$, the velocity $v=(v_1(x,t),\ldots,v_n(x,t))$, and the temperature $\theta=\theta(x,t)$. The internal energy per unit mass, and the entropy per unit mass, are denoted $\epsilon=\epsilon(x,t)$ and $\eta=\eta(x,t)$. The field equations included are the balance of mass, of momentum and (internal) energy, so that
\begin{subequations}\label{eq:fluid:gen}
\begin{equation}\label{eq:fluid:gen:dens}
\Pi^{\rho}:~\dfrac{\partial \rho}{\partial t}  + \dfrac{\partial \rho v_i}{\partial x_i}=0,
\end{equation}
\begin{equation}\label{eq:fluid:gen:mom}
\Pi_i^{v}:~\rho D_t v_i - \dfrac{\partial  T_{ij}}{\partial x_j}- \rho g_i =0,\qquad i=1\ldots, n,\\[2ex]
\end{equation}
\begin{equation}\label{eq:fluid:gen:en}
\Pi^{\epsilon}:~\rho D_t \epsilon +  \dfrac{\partial q_i}{\partial x_i}  -  T_{ij} \dfrac{\partial  v_i}{\partial x_j} =\rho r,
\end{equation}
\end{subequations}
where $\rho r$ denotes the external energy supply density and $\rho g_i$ is the body force, i.e. gravity. Where convenient, for the sake of brevity, we will denote time and space partial derivatives
\[
\frac{\partial\rho}{\partial t}=\partial_t \rho = \rho_t,\qquad \frac{\partial\rho}{\partial x_j}=\partial_j \rho,
\]
etc., and use the Einstein summation convention where appropriate; the material derivative operator is given by
\beq\label{eq:TotD}
D_t = \frac{\partial}{\partial t} +v_j \frac{\partial}{\partial x_j}.
\eeq
In \eqref{eq:fluid:gen}, $T_{ij}=T_{ij}(x,t)$ denotes the fluid stress tensor, and $q_i=q_i(x,t)$, $i=1,\ldots, n$ represent the components of the outgoing energy flux $q(x,t)$. For certain classes of models given by \eqref{eq:fluid:gen}, $\rho$, $v$ and $\theta$ may play the role of dependent variables, and $\epsilon, T_{ij}, q_i$ the role of constitutive functions.

As a basis for the derivation of a set of mathematical constraints on admissible forms of  the constitutive functions \eqref{eq:def:constit} of the given problem, as it will be explained below, one considers an entropy inequality expressing the second law of thermodynamics (SL). For instance, for a medium described by the equations of motion and energy \eqref{eq:fluid:gen}, it may be written in the local form
\begin{equation}\label{eq:entropy_ineq:Liufluid}
\Pi^{\eta}:~\rho D_t\, \eta + \dfrac{\partial \Phi_i}{\partial x_i}  - \rho s \geq 0,\
\end{equation}
where $\eta=\eta(x,t)$ and $s=s(x,t)$ denote respectively the entropy of the medium per unit mass, and the entropy supply (from external sources) per unit mass, and $\Phi = (\Phi_1(x,t),\ldots,\Phi_n(x,t))$ is the outgoing entropy flux vector.

The local entropy inequality \eqref{eq:entropy_ineq:Liufluid} is related with a global formulation of the SL as follows. Consider a material domain $\V(t)$ moving with the fluid within the physical domain $\Omega$. Suppose $\V(t)$ has a piecewise smooth boundary $\p\V(t)$; it is defined  by
\[
\p\V(t) = \left\{ {X}({x},t)~|~ D_t X({x},t)=0\right\},
\]
where $X({x},t)$ are macroscopic fluid particle labels (i.e., material, or Lagrangian coordinates) of the boundary points. The rate of change of the total entropy of the fluid in $\V(t)$ is given by
\beq\label{eq:fluid:entr:vol}
\frac{d}{dt} \int_{\V(t)} \rho \eta \; dV = \int_{\V(t)} \rho s  \; dV  - \oint_{\p\V(t)} \Phi \cdot dA + Q_\V(t).
\eeq

In \eqref{eq:fluid:entr:vol}, $Q_\V(t)$ denotes the additional entropy rate of change within the given volume due to all other physical effects that influence the amount of order in the system. Then the local form of the entropy inequality \eqref{eq:entropy_ineq:Liufluid} provides a sufficient condition that for every material domain $\V(t)$, the additional entropy rate of change $Q_\V(t)$ is non-negative.

As remarked in \cite{hauser2002historical}, in rational thermodynamics, a constitutive model \eqref{eq:def:constit} is considered flawed if there exists a solution of the field equations $\Pi^{\phi}$ that violates the SL.

Condition \eqref{eq:entropy_ineq:Liufluid}, according to \cite{muller_thermodynamic_1968}, has to be interpreted as a restriction on the constitutive functions rather than as restriction on the processes that a body can possibly undergo; in rational thermodynamics, \cite{muller_thermodynamic_1968} proposed a local \emph{entropy inequality} condition.

The requirement of the M\"{u}ller's principle is that the inequality \eqref{eq:entropy_ineq:Liufluid} should hold for every admissible thermodynamic process, i.e., for every solution $\rho, v_i, \theta$ of \eqref{eq:fluid:gen}. This requirement yields restrictions on the constitutive functions \eqref{eq:def:constit}. The specific entropy $\eta$ and the entropy flux $\Phi_i$ are also regarded the unknown constitutive functions.

%
\subsection{M\"{u}ller's Approach}\label{sec:MuellerAppr}

In the context of continuum mechanics, Truesdell \cite{truesdell_mechanical_1962} proposed a set of heuristic postulates on the interdependence and mathematical specification of the distinct phases within a mixture, known as \textit{mixture theory}. Assume that every point in space is simultaneously occupied by every phase, and that each phase is governed by the same balance laws as the mixture, amended by additional terms that account for interchanges between the phases. M\"{u}ller \cite{muller_thermodynamic_1968}, generalizing the previous work of Coleman and Noll \cite{coleman_thermodynamics_1963}, complemented these postulates by suggesting that for every process, i.e. the solutions of the balance laws, the constitutive functions of such a mixture (and its constituents) are restricted by the need to obey the second law of thermodynamics. The latter was posed in terms of an \emph{entropy inequality}, introduced in \cite{muller1967entropy}, expressing the fact that the entropy production in the system is nonnegative.

We now outline the main points of \cite{muller_thermodynamic_1968}, in order to summarize his approach, and compare with the more general and systematic M\"{u}ller-Liu algorithm that will follow.
\begin{enumerate}
  \item A mixture of substances involving several constituents is considered.
  \item The classical PDEs describing the dynamics of the mixture in Eulerian coordinates (balance of mass, momentum, and internal energy, analogs of \eqref{eq:fluid:gen}) are written for each constituent, and for the mixture as a whole. The momentum equations involve the additional source term responsible for inter-species interactions; the energy supply term in the energy equation \eqref{eq:fluid:gen:en} includes a radiation part $\rho r_R$.
  \item An entropy production inequality is written, analogous to \eqref{eq:entropy_ineq:Liufluid} accounting for the entropy advection, fluxes, and an external supply term. It is stated that physically relevant thermodynamic processes satisfy the entropy inequality.
  \item The external entropy supply term in \eqref{eq:entropy_ineq:Liufluid} is assumed to equal the external radiative energy supply divided by the temperature:
  \beq\label{eq:Muller:entr_ener_source}
  s = {r_R}/{\theta}.
  \eeq
  (The condition \eqref{eq:Muller:entr_ener_source} holds exactly for perfect gases, and is not generally true for other substances.)
  \item The constitutive functions of the model, to be determined, include the energy and entropy densities (in terms of the specific free energy function $\psi$) and energy and entropy fluxes (encoded in the flux vectors $k_i$). Additionally, the stress tensor of the medium, the entropy and energy fluxes, and constituent mass production terms are also considered unknown constitutive functions.
  \item In the entropy production inequality, the external entropy supply term $s$ is replaced by the terms of the energy balance equation \eqref{eq:fluid:gen:en} using \eqref{eq:Muller:entr_ener_source}. In the energy balance equation, in turn, the inter-species interaction term is substituted through a similar term in the momentum balance equation. As a result, the  entropy inequality becomes a linear combination involving essential parts of the energy and momentum equations. This form may be referred to as \emph{the extended form of the entropy inequality}.
  \item A form of constitutive functions, involving dependent variables of the problem and their specific partial derivatives, is assumed, and simplified according to Noll's principle of material objectivity.
  \item Several additional simplifying assumptions on the form of constitutive functions are made, including linear dependence on certain higher derivatives of the dependent variables.
  \item The constitutive functions are substituted in the entropy inequality in its extended form. In the result, which is linear in certain highest derivatives of the field variables, the corresponding terms are collected.
  \item Since the extended entropy production inequality is required to hold for \emph{all} solutions of the dynamic balance equations, the independent partial derivatives can assume any value. This allows one to set the corresponding coefficients to zero.

  \item One consequently obtains an underdetermined set of partial differential equations on the unknown constitutive functions, providing restrictions on the previously posed forms of constitutive functions.

\end{enumerate}

%
\subsection{The Liu Algorithm}\label{sec:Mul-Liu}

Müller's procedure (Section \ref{sec:MuellerAppr}) has been modified and generalized by Liu \cite{liu_method_1972,liu2002shih} through the consideration of a constrained entropy inequality and the use of Lagrange multipliers, leading to what is generally referred to as \emph{the entropy principle of Müller and Liu}, or \emph{the Liu algorithm}. In this approach, external supply terms are neglected; it is argued that they do not affect the material behavior. Instead of a sequence of substitutions that yields a linear combination of source-free energy and momentum equations within the entropy inequality, the Müller-Liu procedure yields a similar extended entropy inequality by adding to it a linear combination of the dynamic equations of the model. The application of Liu's lemma to the extended entropy inequality yields constraints on constitutive function forms. Even though formulated for algebraic equations, Lemma \ref{lem:Liu} is traditionally used to analyze entropy-type inequalities, provided that the model of interest is linear in some parametric derivatives.

The Liu algorithm can be utilized for models that do not necessarily have the ``entropy" defined. Instead, one generally considers an inequality formulated for a scalar, additive and objective thermodynamic constitutive quantity $\eta$, which we still refer to as ``entropy" below \cite{hauser2002historical}.

%
%
%

We now outline the main steps of the Liu algorithm, following the works of Liu \cite{liu_method_1972,liu2002shih} and Hausner \& Kirchner \cite{hauser2002historical}, and illustrate them using the physical example of Section \ref{sec:RunEg}. The notation of Section \ref{sec:ConstitModProb} is used.


\begin{enumerate}
  \item For a given physical model, define the fields of interest $\phi$, and the dynamic PDEs $\Pi^{\phi}$. In the running example, we have the dependent variables
\[
\phi = (\rho, v, \theta),
\]
and the governing equations \eqref{eq:fluid:gen} with zero source terms ($g_i, r=0$).

  \item Define the entropy inequality $\Pi^{\eta}$. For our example, it is given by \eqref{eq:entropy_ineq:Liufluid} with zero source terms ($s=0$).

  \item For each of the governing scalar PDEs $\Pi^{\phi}$, define a scalar Lagrange multiplier $\Lambda^{\phi}$. Write down the extended entropy inequality
  \beq\label{eq:Liu:Extended:entropy:ineq}
  \widetilde{\Pi}^{\eta}:~\Pi^{\eta} - \Lambda^{\phi} \Pi^{\phi} \geq 0.
  \eeq
  In the running example, $\widetilde{\Pi}^{\eta}$ is given by
  \beq\label{eq:Liu:Extended:entropy:ineq:RunEx}
  \barr
  \Pi^{\eta} - \Lambda^{\rho} \Pi^{\rho} - \Lambda_i^{v} \Pi_i^{v} - \Lambda^{\epsilon} \Pi^{\epsilon}\\[2ex]
~~= \rho \partial_t \eta +\rho v_i \partial_i \eta+ \partial_i \Phi_i  \\[2ex]
~~~~-\Lambda^{\rho} \left(\partial_t \rho+ \partial_i(\partial \rho v_i) \right)
-\Lambda_i^{v} \left( \rho \,\partial_t v_i +\rho v_j \partial_j v_i  - \partial_j T_{ij} \right)\\[2ex]
~~~~- \Lambda^{\epsilon} \left(\rho\, \partial_t \epsilon +\rho v_i \partial_i \epsilon  +\partial_i q_i -  T_{ij} \partial_j v_i \right)~\geq 0.
\earr
  \eeq

  \item Define the full set of constitutive functions $\psi$ \eqref{eq:def:constit} of the model. This set includes the natural (physical) constitutive functions, additional entropy-related constitutive functions (entropy density and fluxes), and the multipliers $\Lambda^{\phi}$. In the example, following \cite{liu2002shih} Sections 7.2, 7.3, we let
  \beq\label{eq:simplefluid:constit}
  \psi=(\epsilon, q_i, T_{ij}, \eta, \Phi_{i},  \Lambda^{\rho}, \Lambda_i^{v}, \Lambda^{\epsilon}).
  \eeq

  \item Postulate the dependence \eqref{eq:def:constit} on the local variables, including certain field variables, as well as, possibly their derivatives, and independent variables. (Depending on the application, there may be a substantial freedom of choice of the fields to be placed into $\psi$ and $\phi_C$.)  In the running example, we choose a simplified ansatz
  \beq\label{eq:simplefluid:constit:dependence}
  \phi_C=(\rho, \theta, \partial_i \theta).
  \eeq

  \item Substitute the chosen constitutive function forms \eqref{eq:def:constit} into the extended entropy inequality $\widetilde{\Pi}^{\eta}$, carrying out the chain rule of differentiation for every constitutive function:
\beq\label{eq:derivatives:chain}
\frac{\partial \psi}{\partial t}= \sum_{\phi_{C}} \frac{\partial \psi}{\partial \phi_{C}} \frac{\partial \phi_{C}}{\partial t}, \qquad
\frac{\partial \psi}{\partial x_i}= \sum_{\phi_{C}} \frac{\partial \psi}{\partial \phi_{C}} \frac{\partial \phi_{C}}{\partial x_i}.
\eeq
Denote the obtained extended entropy inequality $\widetilde{\Pi}^{\eta}_{(1)}$.

  \item Observe that the extended entropy inequality $\widetilde{\Pi}^{\eta}_{(1)}$ is linear with respect to a set of higher derivatives of the field variables $\phi$. Take the widest set of such derivatives, excluding ones present in the constitutive dependencies $\phi_C$. In the running example, this set of ``arbitrary elements" is given by
   \beq\label{eq:Arbitrary_Elements:Liu:fluid:1}
    Y = ( \partial_t \rho, \partial_i \rho, \partial_t v_i, \partial_j v_i, \partial_t \theta, \partial_{t,i} \theta, \partial_{i,j} \theta).
   \eeq

  \item Collect terms at $\widetilde{\Pi}^{\eta}_{(1)}$ with respect to the set $Y$ of higher derivatives.  In the example, one obtains
  \beq\label{eq:Liu:Extended:entropy:Pi1:collected:RunEx}
  \barr
  \widetilde{\Pi}^{\eta}_{(1)}:~&\dfrac{\partial \rho}{\partial t} \left(\rho \dfrac{\partial \eta}{\partial \rho} - \rho \Lambda^{\epsilon} \dfrac{\partial \epsilon}{\partial \rho} - \Lambda^{\rho}\right) \\[2ex]

&+\dfrac{\partial  \rho}{\partial x_i} \left( \rho v_i \dfrac{\partial  \eta}{\partial \rho}- \Lambda^{\epsilon}\rho v_i  \dfrac{\partial \epsilon}{\partial \rho} -  \Lambda^{\epsilon}\dfrac{\partial q_i}{\partial \rho}+ \Lambda_j^{v} \dfrac{\partial  T_{ij}}{\partial \rho} + \dfrac{\partial \Phi_i}{\partial \rho}- \Lambda^{\rho} v_i \right)\\[2ex]

&-\dfrac{\partial v_i}{\partial t}  \Lambda_i^{v} \rho +\dfrac{\partial v_i}{\partial x_j} \left(\Lambda^{\epsilon}  T_{ij}  -  \Lambda_i^{v} v_j \rho - \Lambda^{\rho}  \rho \delta_{ij} \right) \\[2ex]

&+\rho \dfrac{\partial \theta}{\partial t} \left(\dfrac{\partial \eta}{\partial \theta} -\Lambda^{\epsilon} \dfrac{\partial \epsilon}{\partial \theta} \right)\\[2ex]

&+\rho\dfrac{\partial^2 \theta}{\partial t \,\partial x_i} \left(\dfrac{\partial \eta}{\partial (\partial_i \theta)} -\Lambda^{\epsilon} \dfrac{\partial \epsilon}{\partial (\partial_i \theta)} \right)   \\[2ex]

&+\dfrac{\partial^2  \theta}{\partial x_i \partial x_j} \left( \rho  v_j \dfrac{\partial  \eta}{\partial (\partial_i \theta)}- \Lambda^{\epsilon}\rho  v_j \dfrac{\partial \epsilon}{\partial (\partial_i \theta)} -  \Lambda^{\epsilon}\dfrac{\partial q_j}{\partial (\partial_i \theta)}+ \Lambda_k^{v} \dfrac{\partial  T_{jk}}{\partial (\partial_i \theta)} + \dfrac{\partial \Phi_j}{\partial (\partial_i \theta)}  \right)\\[2ex]

&+\dfrac{\partial  \theta}{\partial x_i} \left( \rho v_i  \dfrac{\partial  \eta}{\partial \theta}- \Lambda^{\epsilon}\rho  v_i \dfrac{\partial \epsilon}{\partial\theta} -  \Lambda^{\epsilon}\dfrac{\partial q_i}{\partial \theta}+ \Lambda_j^{v} \dfrac{\partial  T_{ij}}{\partial \theta}+\dfrac{\partial \Phi_i}{\partial \theta}  \right) \geq 0.
\earr
\eeq


  \item In the spirit of Lemma \ref{lem:Liu}, set to zero coefficients at the arbitrary elements $Y$. Obtain a set of \emph{Liu identities} and a \emph{residual inequality}. For the running example, the Liu identities are given by
\beq\label{eq:Liu:RunEx:LiuID}
\barr
\rho \dfrac{\partial \eta}{\partial \rho} - \rho \Lambda^{\epsilon} \dfrac{\partial \epsilon}{\partial \rho} - \Lambda^{\rho}=0, \\[2ex]
\rho v_i \dfrac{\partial  \eta}{\partial \rho}- \Lambda^{\epsilon}\rho v_i  \dfrac{\partial \epsilon}{\partial \rho} -  \Lambda^{\epsilon}\dfrac{\partial q_i}{\partial \rho}+ \Lambda_j^{v} \dfrac{\partial  T_{ij}}{\partial \rho} + \dfrac{\partial \Phi_i}{\partial \rho}- \Lambda^{\rho} v_i =0;\\[3ex]

\Lambda_i^{v} \rho=0, \qquad \Lambda^{\epsilon}  T_{ij}  -  \Lambda_i^{v} v_j \rho - \Lambda^{\rho}  \rho \delta_{ij} =0,\\[2ex]

\dfrac{\partial \eta}{\partial \theta} -\Lambda^{\epsilon} \dfrac{\partial \epsilon}{\partial \theta} =0,\qquad \dfrac{\partial \eta}{\partial (\partial_i \theta)} -\Lambda^{\epsilon} \dfrac{\partial \epsilon}{\partial (\partial_i \theta)} =0,\\[2ex]

\rho  v_j \dfrac{\partial  \eta}{\partial (\partial_i \theta)}- \Lambda^{\epsilon}\rho  v_j \dfrac{\partial \epsilon}{\partial (\partial_i \theta)} -  \Lambda^{\epsilon}\dfrac{\partial q_j}{\partial (\partial_i \theta)}+ \Lambda_k^{v} \dfrac{\partial  T_{jk}}{\partial (\partial_i \theta)} + \dfrac{\partial \Phi_j}{\partial (\partial_i \theta)} =0,\\[2ex]
\earr
\eeq
and the residual inequality is given by
\beq\label{eq:Liu:RunEx:ResIneq}
\dfrac{\partial  \theta}{\partial x_i} \left( \rho v_i  \dfrac{\partial  \eta}{\partial \theta}- \Lambda^{\epsilon}\rho  v_i \dfrac{\partial \epsilon}{\partial\theta} -  \Lambda^{\epsilon}\dfrac{\partial q_i}{\partial \theta}+ \Lambda_j^{v} \dfrac{\partial  T_{ij}}{\partial \theta}+\dfrac{\partial \Phi_i}{\partial \theta}  \right) \geq 0.
\eeq

  \item Solve the Liu identities to obtain constraints on the constitutive functions. For the example of a simple fluid, one obtains a set of constraints consistent with those presented in \cite{liu2002shih}. In particular, one has the following.
  \begin{itemize}
    \item The requirement of isotropy: $T_{ij}=-\delta_{ij} p$, where $p=p(\rho,\theta)$ is the hydrostatic pressure.
    \item The vanishing Lagrange multipliers $\Lambda_i^{v}=0$.
    \item The classical form of the energy Lagrange multiplier $\Lambda^{\epsilon}=F(\theta)$. 
    In particular, in many works including that of M\"{u}ller, one has $\Lambda^{\epsilon}=1/\theta$, see e.g. \cite{muller_thermodynamics_1984}.

    \item The form of the entropy mass density:
    \beq\label{eq:fluid:Liu:entr:finalform}
    \eta = \dfrac{F^2}{F'}\int \dfrac{p}{\rho^2}\, d\rho + K,
    \eeq
    where $F=F(\theta)$, $K=K(\theta)$ are arbitrary functions.
    \item The form of the energy mass density:
    \beq\label{eq:fluid:Liu:e:finalform}
    \epsilon = \dfrac{1}{F'}\int \dfrac{(pF)_\theta}{\rho^2}\, d\rho.
    \eeq
    \item Relationships between for energy and entropy fluxes, which do depending on the components of the temperature gradient.
  \end{itemize}
  Here the functions $P,F,G$ are arbitrary functions of the indicated arguments.

\end{enumerate}


%

\section{Constitutive Modelling using the Liu Algorithm: a Symbolic Implementation}\label{sec:Maple}

The symbolic software package \verb|GeM| for \verb|Maple| contains routines for the computation of local conservation laws and Lie point and higher-order local symmetries of ordinary and partial differential equations and  ODE/PDE systems \cite{cheviakov2007gem, cheviakov2010symbolic,GemReferenceOnline}. The package can also be efficiently used for other computations that involve symbolic manipulation of differential equations (DE), their differential consequences, and related expressions involving independent and dependent variables, partial derivatives, arbitrary (constitutive) parameters, and arbitrary (constitutive) functions. In particular, equivalence transformations of DE families can be studied \cite{cheviakov2017symbolic}. In conjunction with the standard \verb|Maple| routine \verb|rifsimp| for the reduction of overdetermined systems and case splitting, the \verb|GeM| package has been successfully applied to many previously intractable problems of symmetry and conservation law analysis and classification.

The routines of the \verb|GeM| package use a computationally efficient representation of differential equations, through the conversion of the dependent variables and their derivatives to \verb|Maple| symbols instead of functions or expressions. Then the chain rule for differential functions of dependent variables is simplified to standard partial derivative operations. For example, the
\verb|Maple| representation of a function $F(x,y,z)$ and its partial derivative by $F_x$ in \verb|GeM| routines is respectively $\verb|F|$ and $\verb|Fx|$. Consequently, for constitutive functions that depend on functions (dependent variables and their derivatives), the differentiations significantly simplify; e.g., if $H=H\left(F(x,y,z), F_x(x,y,z)\right)$, then in the \verb|GeM| representation, one has $\verb|H=H(F, Fx)|$. Its partial derivative, for example by $y$, is represented as
\[
\dfrac{\partial}{\partial y} H\left(F(x,y,z), F_x(x,y,z)\right) = \verb|diff(H(F, Fx), F) * Fy + diff(H(F, Fx), Fx) * Fxy|,
\]
and does not involve functions of functions. For details of data representation, routines, options, methods, and examples, see \cite{cheviakov2007gem, cheviakov2010symbolic}.

In the current work, the \verb|GeM| package is employed to automate computations within the Liu algorithm (Section \ref{sec:Mul-Liu}). The sequence of steps is outlined below, starting from an elementary example of one-dimensional gas flow, and continuing with the heat conducting compressible anisotropic fluid (see Section \ref{sec:RunEg}).
%
\subsection{Symbolic Example 1: One-Dimensional Gas Dynamics Equations}\label{sec:Maple:EG1}

As a first example, we consider a one-dimensional compressible gas flow, following the notation of \cite{grossman2000fundamental}. The continuity, momentum, and energy equations, in the absence of external supply terms, are given by
\begin{subequations}\label{eq:gas1D:gen}
\begin{equation}\label{eq:gas1D:gen:dens}
\Pi^{\rho}:~\rho_t  + (\rho v)_x=0,
\end{equation}
\begin{equation}\label{eq:gas1D:gen:mom}
\Pi^{v}:~\rho (v_t+vv_x) + p_x =0, \\[2ex]
\end{equation}
\begin{equation}\label{eq:gas1D:gen:en}
\Pi^{\epsilon}:~\rho (\epsilon_t+ v \epsilon_x) + pv_x =0,
\end{equation}
\end{subequations}
where $\rho(t,x)$ is the mass density, $v(t,x)$ is the scalar spatial velocity in the $x$-direction, $p(t,x)$ is the scalar pressure, and $\epsilon(t,x)$ is the thermal energy per unit mass. An additional ``internal" dependent variable not explicitly present in the model is the temperature $\theta(t,x)$.

The model \eqref{eq:gas1D:gen} involves four unknowns and three equations, and thus requires a closure, a constitutive relationship, which will be determined through an entropy principle. For simplicity, let us assume the gas to be calorically perfect, see \cite{grossman2000fundamental}, that is, satisfying $\epsilon=C_v\theta$, and $C_v=\const$ is the specific heat at constant volume. We also suppose that the gas is thermally perfect, i.e., satisfies the ideal gas law,
\beq\label{eq:1Dgas:ideal}
p=\rho \tilde{R} \theta,
\eeq
where $\tilde{R}=R/M$ is the specific gas constant, and $R$ and $M$ are the universal gas constant and the molar mass of the gas. We also note the relationship
\[
C_v = \dfrac{i}{2}\tilde{R} = \dfrac{\tilde{R}}{\gamma-1},
\]
where $i$ is the number of degrees of freedom of a gas molecule, and $\gamma=(i+2)/i$ is the adiabatic exponent.

Following the Liu algorithm, we define the entropy inequality by
\begin{equation}\label{eq:entropy_ineq:1Dgas}
\Pi^{\eta}:~\rho (\eta_t+v\eta_x) \geq 0,
\end{equation}
with $\eta(t,x)$ denoting the mass density of entropy. As seen in \cite{grossman2000fundamental,muller1998rational}, for an inviscid non-heat conducting substance, the external supply terms in the energy and entropy equations are proportional, so it is valid to assume that they vanish simultaneously.
The extended entropy inequality is given by
\beq\label{eq:Liu:Extended:entropy:ineq:1Dgas}
\widetilde{\Pi}^{\eta}:~\Pi^{\eta} - \Lambda^{\rho} \Pi^{\rho} - \Lambda^{v} \Pi^{v} - \Lambda^{\epsilon} \Pi^{\epsilon} \geq 0.
\eeq
We let $\rho$, $v$, $\epsilon$ play the role of dependent variables, and allow the five constitutive functions \eqref{eq:simplefluid:constit}
\beq\label{eq:gas1d:constit}
\psi=(p, \eta, \Lambda^{\rho},  \Lambda^{v}, \Lambda^{\epsilon})
\eeq
to depend on
\beq\label{eq:constit:dependence:1Dgas}
\phi_C=(\rho, \epsilon).
\eeq

%
%
%

The computation following the Liu algorithm proceed as follows.

\medskip\noindent \textbf{Step A. Initialize.} Clear the variables. Initialize the package.
\beq\label{eq:gem:initeq}
\barr
\verb|restart:|\\
\verb|read("d:/gem32_12.mpl"):|\\
\earr
\eeq

\medskip\noindent \textbf{Step B. Declare variables and constitutive functions.}
\[
\barr
\verb|ind:=t,x;   dep:=R(ind), V(ind), E(ind);|\\
\verb|Constit_Dependence:=R(ind), E(ind);|\\
\verb|Constit_F:=P(Constit_Dependence), S(Constit_Dependence),|\\
\verb|           LR(Constit_Dependence), LV(Constit_Dependence), |\\
\verb|           LE(Constit_Dependence) ;|
\earr
\]
\[
\barr
\verb|gem_decl_vars(indeps=[ind], deps=[dep], freefunc=[Constit_F]);|
\earr
\]

Here we used the \verb|Maple| notation $\rho=\verb|R|$, $p=\verb|P|$, $v=\verb|V|$,  $\epsilon=\verb|E|$, $\eta=\verb|S|$ for the fields, and $\Lambda^{\rho}=\verb|LR|$ $\Lambda^{v}=\verb|LV|$, $\Lambda^{\epsilon}=\verb|LE|$ for the Lagrange multipliers. It is our common convention to use small letters for independent variables and capitals for dependent variables and constitutive functions.

\medskip\noindent \textbf{Step C. Declare the model equations.} The PDEs \eqref{eq:gas1D:gen} are defined as follows.
\[
\barr
\verb|Pi_Rho:=diff(R(t, x), t) + diff(R(t, x)*V(t, x), x) = 0;|\\
\verb|Pi_V:=R(t, x)*(diff(V(t, x), t) + V(t, x)*diff(V(t, x), x) |\\
\verb|          + diff(P(Constit_Dependence),x)= 0;|\\
\verb|Pi_E:=R(t, x)*(diff(E(t, x), t) + V(t, x)*diff(E(t, x), x) |\\
\verb|          + P(Constit_Dependence)*diff(V(t, x),x)= 0;|\\
\earr
\]
\[
\barr
\verb|gem_decl_eqs( [Pi_Rho, Pi_V, Pi_E] );|
\earr
\]
The last command declares the given PDEs in terms of \verb|Maple| symbols. The symbolic representation of the left-hand sides of the PDEs (without ``$=0$") is stored in the internal variables \verb|GEM_ALL_EQ_AN|; they can be extracted as follows:
\[
\barr
\verb|Eq_R_Symb:=GEM_ALL_EQ_AN[1];|\\
\verb|Eq_U_Symb:=GEM_ALL_EQ_AN[2];|\\
\verb|Eq_E_Symb:=GEM_ALL_EQ_AN[3];|
\earr
\]
For example, the density equation has the form $\verb|Eq_R_Symb = Rt+R*Vx+Rx*V|$, with all derivatives of dependent variables replaced by \verb|Maple| symbols.

\medskip\noindent \textbf{Step D. The entropy inequality; the extended entropy inequality.} The left-hand side of the entropy inequality \eqref{eq:entropy_ineq:1Dgas} is defined as
\[
\barr
\verb|Pi_S:= R(ind)*diff(S(Constit_Dependence), t)|\\
\verb|      + R(ind)*V(ind)*diff(S(Constit_Dependence), x)|\\
\verb|      + diff(Phi1(Constit_Dependence),x);|
\earr
\]
It is subsequently converted into a \verb|Maple| expression using the \verb|GeM| command
\[
\barr
\verb|Eq_S_Symb:=gem_analyze(Pi_S);|
\earr
\]
The left-hand side of the extended entropy inequality \eqref{eq:Liu:Extended:entropy:ineq:1Dgas} is obtained:
\[
\barr
\verb|Entropy_Ineqality:= Eq_S_Symb - LR(R, E)* Eq_R_Symb| \\
\verb|                     - LV(R, E)* Eq_U_Symb - LE(R, E)* Eq_E_Symb;|
\earr
\]
this expression is linear in terms of the higher derivatives
\[
Y= (\rho_t, \rho_x, v_t, v_x, \epsilon_t, \epsilon_x),
\]
which are not parts of the constitutive dependence $\phi_C$ \eqref{eq:constit:dependence:1Dgas}.

\medskip\noindent \textbf{Step E. Obtain a system of constraints.} According to the Liu's lemma, one now sets to zero coefficients at elements of $Y$, and obtains a split set of constraints, and a residual entropy inequality.

The independent terms in the extended entropy inequality can be collected as follows:
\[
\barr
\verb|Y:= [ Rt, Rx, Vt, Vx, Et, Ex ];|\\
\verb|collect(Entropy_Ineqality, Y);|
\earr
\]
the output is
\[
\barr
\verb|((diff(S(R, E), R))*R-LR(R, E))*Rt|\\
\verb|+((diff(S(R, E), R))*R*V-LR(R, E)*V-LV(R, E)*(diff(P(R, E), R)))*Rx|\\
\verb|-LV(R, E)*Vt*R+(-LV(R, E)*R*V-LR(R, E)*R-LE(R, E)*P(R, E))*Vx|\\
\verb|+((diff(S(R, E), E))*R-LE(R, E)*R)*Et|\\
\verb|+((diff(S(R, E), E))*R*V-LE(R, E)*R*V-LV(R, E)*(diff(P(R, E), E)))*Ex|
\earr
\]
The \verb|collect| step is not necessary -- it simply formats the extended entropy inequality as a sum of terms proportional to the elements of $Y$. Moreover, one observes that the residual entropy inequality is zero, since every term in the output above is proportional to an arbitrary element contained in $Y$.

The actual split set of constraints (Liu identities) is obtained by setting to zero the coefficients of the elements of $Y$, as follows:
\[
\barr
\verb|coeffs_constraints:=[coeffs(Entropy_Ineqality, Y)];|
\earr
\]
The six corresponding constraints are given by
\[
\barr
 \rho\Lambda^{\rho} +  \rho v \Lambda^{v} + p\Lambda^{\epsilon}=0;&
 \rho v \dfrac{\partial \eta}{\partial \epsilon} - \Lambda^{v}\dfrac{\partial p}{\partial \epsilon} - \rho v \Lambda^{\epsilon}=0;\\[2ex]
 \rho \dfrac{\partial \eta}{\partial \rho} - \Lambda^{\rho}=0;&
 v\left(\rho \dfrac{\partial \eta}{\partial \rho} - \Lambda^{\rho}\right) - \Lambda^{v}\dfrac{\partial p}{\partial \rho}=0;\\[3ex]
 \rho \Lambda^{v}=0;&
 \rho \left(\dfrac{\partial \eta}{\partial \epsilon} - \Lambda^{\epsilon}\right)=0.
\earr
\]

\bigskip\noindent \textbf{Step F. Rif-simplify the constraints.}
At this optional step, one can use the \verb|Maple| \verb|rifsimp| routine for the Gr\"obner basis-based reduction of the overdetermined linear system of determining equations obtained at the previous step.
\[
\barr
\verb|all_f := [P(R, E), S(R, E), LR(R, E), LV(R, E), LE(R, E)];|\\
\verb|simplified_eqs := DEtools[rifsimp](coeffs_constraints, all_f, mindim = 1);|
\earr
\]
This  step is particularly important for large systems of constraint equations, since through the elimination of the redundancy, it leads to a significant simplification and reduction of the number of PDEs in the set of constraints. The output of \verb|rifsimp| is a \verb|Maple| table. The simplified constraints are stored in \verb|simplified_eqs[Solved]|.

When used with \verb|mindim = 1| option, the \verb|rifsimp| routine determines the dimension of solution space without solving the DEs: \verb|simplified_eqs[dimension]|. The infinity value corresponds to the presence of arbitrary functions.
Another option, \verb|casesplit|, may be also useful in specific settings, when the most general situation needs to be avoided (such as in the case of an \emph{a priori} vanishing constitutive function, etc.; see \cite{cheviakov2007gem,cheviakov2010symbolic,GemReferenceOnline} and \verb|Maple| help for details).

In our case, the \verb|rifsimp| output \verb|simplified_eqs[Solved]| contains a simple set of constraints
\beq\label{eq:gas1d:rifsimped}
\barr
\Lambda^{v}=0, \qquad \dfrac{\partial \eta}{\partial \rho} = \dfrac{\Lambda^{\rho}}{\rho},\qquad\dfrac{\partial \Lambda^{\epsilon}}{\partial \rho} = \dfrac{1}{\rho}\dfrac{\partial \Lambda^{\rho}}{\partial \epsilon},\qquad \dfrac{\partial \eta}{\partial \epsilon}=\Lambda^{\epsilon},\qquad p = - \rho \dfrac{\Lambda^{\rho}}{\Lambda^{\epsilon}}.
\\[2ex]
\earr
\eeq

\bigskip\noindent \textbf{Step G. Solve the Liu identities.} The PDEs \eqref{eq:gas1d:rifsimped} can be readily solved by hand, or using the built-in \verb|Maple| PDE solver:
\[
\verb|pdsolve(simplified_eqs[Solved],all_f);|
\]
The resulting solution is given by
\beq\label{eq:gas1d:solution}
\barr
\Lambda^{\rho}(\rho, \epsilon)=\displaystyle\rho \int F_\rho\,d \epsilon + G(\rho),\quad \Lambda^{v}(\rho, \epsilon)=0,\quad \Lambda^{\epsilon}(\rho, \epsilon)=F(\rho, \epsilon),\\[2ex]
p(\rho, \epsilon)=- \rho \dfrac{\Lambda^{\rho}}{\Lambda^{\epsilon}},\quad \eta=\displaystyle \int F\,d \epsilon + \displaystyle \int \dfrac{G}{\rho}\,d \rho,
\earr
\eeq
where $F(\rho, \epsilon)$ and $G(\rho)$ are arbitrary functions. Alternatively, one may treat $p=p(\rho, \epsilon)$ as an arbitrary function, and express the restrictions on $\eta$ and the multipliers in its terms.
The solution \eqref{eq:gas1d:solution} is in agreement with \cite{muller1998rational}. Moreover,  PDEs \eqref{eq:gas1d:rifsimped} yield a well-known formula
\beq\label{eq:gas1d:solution:p:eta}
p=-\rho^2\dfrac{\eta_\rho}{\eta_\epsilon}.
\eeq
It is easy to verify that indeed, for any $\eta(\rho, \epsilon)$ satisfying \eqref{eq:gas1d:solution:p:eta}, the original entropy inequality
$\Pi^{\eta}$ \eqref{eq:entropy_ineq:1Dgas} vanishes identically as long as \eqref{eq:gas1D:gen:dens}, \eqref{eq:gas1D:gen:en} are satisfied, that is, on all solutions of the given model.

For the case of an ideal gas, with
\[
p=\rho \hat{R} \,\theta, \quad \eta = C_v \ln (\rho^{1-\gamma}\,\theta ),
\]
one has a specific instance of the solution \eqref{eq:gas1d:solution}:
\[
\barr
p(\rho, \epsilon)=(\gamma-1) \rho \epsilon,\quad \eta(\rho, \epsilon) = C_v \ln \left(\dfrac{\epsilon}{C_v\, \rho^{\gamma-1}}\right),\\[3ex]
\Lambda^{\rho}(\rho, \epsilon) = - \hat{R}, \quad \Lambda^{v}(\rho, \epsilon)=0,\quad \Lambda^{\epsilon}(\rho, \epsilon)=\dfrac{C_v}{\epsilon}=\dfrac{1}{\theta}.
\earr
\]
In particular, the form of $\Lambda^{\epsilon}$ is a well-known result, commonly arising in constitutive modelling with Lagrange multipliers.
%
%
\subsection{Symbolic Example 2: Heat-conducting anisotropic fluid}\label{sec:Maple:EG2}

We now apply the same \verb|Maple| algorithm to the running example of a two-dimensional ($n=2$) simple heat conducting compressible anisotropic fluid (cf. Section \ref{sec:RunEg}). The computations presented here are the ones leading to the formulas discussed above in Section \ref{sec:Mul-Liu}.

\medskip\noindent \textbf{Step A. Initialize.} The initialization proceeds using \eqref{eq:gem:initeq}.

\medskip\noindent \textbf{Step B. Declare variables and constitutive functions.}
\[
\barr
\verb|ind:=t,x,y;   dep:=R(ind), U(ind), V(ind), W(ind);|\\
\verb|gem_decl_vars(indeps=[ind], deps=[dep]);|
\earr
\]
The independent variables are given by $t,x,y$; in \verb|Maple| notation, the physical fields are the density $\rho=\verb|R|$, the velocity in the $x$-direction $u=\verb|U|$, the velocity in the $y$-direction $v=\verb|V|$, the temperature $\theta=\verb|W|$. Unlike the previous example, here the temperature is considered a field, while the internal energy $\epsilon$ is treated as a constitutive function. The constitutive dependence $\phi_C$ and the constitutive functions $\psi$ are defined according to \eqref{eq:simplefluid:constit},    \eqref{eq:simplefluid:constit:dependence}, using the commands
\[
\barr
\verb|Constit_Dependence:=R(ind),W(ind),diff(W(ind),x),diff(W(ind),y);|\\
\verb|Constit_F:=T11(Constit_Dependence), T12(Constit_Dependence),|\\
\verb|           T22(Constit_Dependence),|\\
\verb|           Q1(Constit_Dependence), Q2(Constit_Dependence),|\\
\verb|           E(Constit_Dependence), S(Constit_Dependence),|\\
\verb|           Phi1(Constit_Dependence), Phi2(Constit_Dependence),|\\
\verb|           LR(Constit_Dependence), LU(Constit_Dependence), |\\
\verb|           LV(Constit_Dependence), LE(Constit_Dependence) ;|
\earr
\]
Instead of an isotropic pressure, the current example features a symmetric stress tensor, with three independent entries denoted by $T_{11}=\verb|T11|$, $T_{12}=T_{21}=\verb|T12|$, $T_{22}=\verb|T22|$. The heat flux is denoted by $q_{1}=\verb|Q1|$, $q_{2}=\verb|Q2|$, the entropy and the internal energy by $\eta=\verb|S|$, $\epsilon=\verb|E|$, and the entropy flux components by $\Phi_{1}=\verb|Phi1|$, $\Phi_{2}=\verb|Phi2|$. The \verb|Maple| expressions \verb|LR|, \verb|LU|, \verb|LV|, \verb|LE| denote the respective Lagrange multipliers. All these quantities are declared using the command
\[
\barr
\verb|gem_decl_vars(indeps=[ind], deps=[dep], freefunc=[Constit_F]);|
\earr
\]

\medskip\noindent \textbf{Step C. Declare the model equations.}
Before defining the balance equations, we introduce the symbolic material derivative operator \eqref{eq:TotD}:
\[
\barr
\verb|MaterialDer:=Q->diff(Q,t)+U(ind)*diff(Q,x)+V(ind)*diff(Q,y);|
\earr
\]
The PDEs are input as follows.
\[
\barr
\verb|Pi_Rho:=diff(R(ind), t) + diff(R(ind)*U(ind), x) |\\
\verb|        + diff(R(ind)*V(ind), y)= 0;|\\
\verb|Pi_U:=R(ind)*MaterialDer(U(ind)) - diff(T11(Constit_Dependence),x) |\\
\verb|          - diff(T12(Constit_Dependence),y)= 0;|\\
\verb|Pi_V:=R(ind)*MaterialDer(V(ind)) - diff(T12(Constit_Dependence),x) |\\
\verb|          - diff(T22(Constit_Dependence),y)= 0;|\\
\verb|Pi_E:=R(ind)*MaterialDer(E(Constit_Dependence))  |\\
\verb|        +diff(Q1(Constit_Dependence),x) |\\
\verb|        +diff(Q2(Constit_Dependence),y)|\\
\verb|         -T11(Constit_Dependence)*diff(U(ind),x)|\\
\verb|         -T12(Constit_Dependence)*diff(U(ind),y)|\\
\verb|         -T12(Constit_Dependence)*diff(V(ind),x)|\\
\verb|         -T22(Constit_Dependence)*diff(V(ind),y)=0;|\\
\earr
\]
The equations are declared and converted to \verb|Maple| symbolic expressions using
\[
\barr
\verb|gem_decl_eqs( [Pi_Rho, Pi_U, Pi_V, Pi_E] );|
\earr
\]
and are copied into the variables
\[
\barr
\verb|Eq_R_Symb:=GEM_ALL_EQ_AN[1];|\\
\verb|Eq_U_Symb:=GEM_ALL_EQ_AN[2];|\\
\verb|Eq_V_Symb:=GEM_ALL_EQ_AN[3];|\\
\verb|Eq_E_Symb:=GEM_ALL_EQ_AN[4];|
\earr
\]

\medskip\noindent \textbf{Step D. The entropy inequality; the extended entropy inequality.} The left-hand side of the entropy inequality \eqref{eq:entropy_ineq:Liufluid} ($s=0$) is defined and converted into a \verb|Maple| expression using the commands
\[
\barr
\verb|Pi_S:= R(ind)*MaterialDer(S(Constit_Dependence)) +diff(Phi1(Constit_Dependence),x)| \\
\verb|       +diff(Phi2(Constit_Dependence),y);| \\
\verb|Eq_S_Symb:=gem_analyze(Pi_S);|
\earr
\]
The left-hand side of the extended entropy inequality \eqref{eq:Liu:Extended:entropy:ineq:RunEx} is given by
\[
\barr
\verb|Entropy_Ineqality:= Eq_S_Symb - LR(R, W, Wx, Wy)* Eq_R_Symb| \\
\verb|                     - LU(R, W, Wx, Wy)* Eq_U_Symb|\\
\verb|                     - LV(R, W, Wx, Wy)* Eq_V_Symb|\\
\verb|                     - LE(R, W, Wx, Wy)* Eq_E_Symb;|
\earr
\]

\medskip\noindent \textbf{Step E. Obtain a system of constraints.}
The independent terms in the extended entropy inequality can be isolated as follows:
\[
\barr
\verb|Y:= [ Rt, Rx, Ry, Ut, Ux, Uy, Vt, Vx, Vy, Wt, Wxx, Wxy, Wyy, Wtx, Wty ];|\\
\verb|collect(Entropy_Ineqality, Y);|
\earr
\]
The split set of constraints, i.e., the  Liu identities, is obtained as follows:
\[
\barr
\verb|coeffs_constraints:=[coeffs(Entropy_Ineqality, Y)];|
\earr
\]
It is given by the formulas \eqref{eq:Liu:RunEx:LiuID}.
%
%
%
%
%
%

\bigskip\noindent \textbf{Step F. Rif-simplify the constraints.} The simplification step proceeds the same way as for the gas dynamics example. This time, following \cite{liu_method_1972} we apply additional constraints in order to simplify the computations.
In particular, we require that the Lagrange multiplier of the energy equation does not vanish, $\Lambda^{\epsilon} \neq 0$, and both the internal energy and entropy are functions of the temperature but not of the temperature gradient: i.e. $\eta = \eta (\theta,\rho)$ and $\epsilon = \epsilon(\epsilon,\rho)$. In order to achieve these restrictions, we append to the determining equations the conditions
\[
\dfrac{\partial \eta}{\partial \theta_x}=\dfrac{\partial \eta}{\partial \theta_y}=\dfrac{\partial \epsilon}{\partial \theta_x}=\dfrac{\partial \epsilon}{\partial \theta_y}=0,\qquad \Lambda^\epsilon \ne 0.
\]
The
\[
\barr
\verb|all_f := [T11(R, W, Wx, Wy), T12(R, W, Wx, Wy), T22(R, W, Wx, Wy),|\\
\verb|          Q1(R, W, Wx, Wy), Q2(R, W, Wx, Wy),|\\
\verb|          Phi1(R, W, Wx, Wy), Phi2(R, W, Wx, Wy), |\\
\verb|          E(R, W, Wx, Wy), S(R, W, Wx, Wy),|\\
\verb|          LR(R, W, Wx, Wy), LU(R, W, Wx, Wy),|\\
\verb|          LV(R, W, Wx, Wy), LE(R, W, Wx, Wy)];|\\
\verb|simplified_eqs := DEtools[rifsimp]([coeffs_constraints[], |\\
\verb| LE(R, W, Wx, Wy)<>0, |\\
\verb| diff(S(R, W, Wx, Wy),Wx)=0, diff(S(R, W, Wx, Wy),Wy)=0,|\\
\verb| diff(E(R, W, Wx, Wy),Wx)=0, diff(E(R, W, Wx, Wy),Wy)=0,|\\
\verb| diff(S(R, W, Wx, Wy),W)<>0, diff(E(R, W, Wx, Wy),W)<>0],|\\
\verb|all_f, mindim=1);|\\
\verb|simplified_eqs[Solved];|
\earr
\]

\bigskip\noindent \textbf{Step G. Solve the Liu identities.} The PDEs \eqref{eq:gas1d:rifsimped} are subsequently solved, for example, symbolically:
\[
\verb|pdsolve(simplified_eqs[Solved],all_f);|
\]
The output contains the results discussed in the end of Section \ref{sec:Mul-Liu}. In particular, The Lagrange multipliers have the form
\[
\Lambda_{u}=\Lambda_{v}=0, \qquad \Lambda_{\rho}=H,\qquad \Lambda_{\epsilon}=\dfrac{1}{G}
\]
in terms of the arbitrary functions $G=G(\theta)$ and $H=H(\rho,\theta)$. For the stress components, one has
\[
T_{12}=0,\qquad T_{11}=T_{22}=\rho\, G\, H = - p,
\]
that is, the requirement of fluid isotropy, and the form of the hydrostatic pressure $p=p(\rho,\theta)$ consistent with the entropy principle. The internal entropy and energy density forms are consequently given by \eqref{eq:fluid:Liu:entr:finalform} and \eqref{eq:fluid:Liu:e:finalform}. The results are therefore in agreement with \cite{liu_method_1972}.

\section{Discussion and Conclusions}\label{sec:discussion}

In the present work, a symbolic software implementation of the entropy principle of  M\"{u}ller and Liu, a tool to derive thermodynamically consistent constraints for the constitutive functions of a system in continuum mechanics, has been presented and discussed. 
Two basic examples are analyzed: a one-dimensional gas dynamics model, and a model of a anisotropic compressible heat conducting fluid. The entropy inequality and the extended entropy inequality were presented as \verb|Maple| expressions in terms of the independent and dependent variables of the problem and their derivatives (treated as \verb|Maple| symbols), and unknown constitutive functions (treated as functions of several simple scalar variables). Using the routines of \verb|GeM| package, the coefficients at independent higher-order derivatives of field variables were automatically set to zero, yielding Liu identities. The latter were efficiently simplified and solved using standard \verb|Maple| commands. The results obtained provided the required restrictions on the constitutive functions; these restrictions are physically meaningful and are in agreement with the literature.

The presented automation of the common Liu algorithm is particularly important since such computations, when done by hand, tend to be technically demanding, time-consuming, and prone to human error. For the examples considered in this work, the full set of computations performed on a standard desktop computer took seconds. The \verb|Maple|-based symbolic framework developed in the current contribution can be applied with minimal changes to treat similar as well as more complicated models; the program sequence can also be naturally modified, if necessary. Nonetheless, the presented scheme is not being universally applicable to a model of arbitrary complexity. Steps A to E can be carried out efficiently for virtually any set of equations and modelling assumptions, however, steps F (symbolic simplification of Liu identities) and G (their symbolic solution) may not be accessible if the dependency $\phi_C$ of the constitutive functions involves a large number of variables. Exact numbers that are symbolically tractable depends on the model; for example, in a similar \verb|Maple|-based computation of local conservation laws of Euler equations \cite{cheviakov2014generalized}, the unknown multipliers depended on 45 variables, including partial derivatives of the fields; however, the conservation law determining equations were substantially more overdetermined. The Liu identities in constitutive modelling are usually less overdetermined, in particular, they never include conditions on the derivatives of the Lagrange multipliers $\Lambda^{\phi}$.

Several research directions aimed at the extension of the symbolic algorithm outlined in this work can be named.
\begin{enumerate}
  \item The presented procedure can be generalized to aid in constitutive modelling for \emph{thermodynamic equilibria}, the states of minimal (i.e., zero) entropy production. In these states, additional equations are formulated that serve as constraints for the equilibrium parts of the unknown constitutive functions \cite[p. 306ff]{hutter_continuum_2004}, based on the residual entropy inequality.

  \item Depending on the model, the solution of Liu identities may not lead to one set of closed-form expressions for constitutive functions. In related symbolic computations, the \verb|rifsimp| routine used without the \verb|casesplit| option, as done in the above example, returns the simplified equations in the most general case, when no pivot coefficient is zero. Pivoting based on zero or nonzero values of certain coefficients may lead to different cases, for example, more general forms of some constitutive functions and less general forms of the other ones. The \verb|casesplit| option of the \verb|Maple| \verb|rifsimp| routine may be useful to produce investigate \emph{trees} of such special cases. Indeed, in specific physical situations, assumptions that some coefficient (such as a Lagrange multiplier or its certain partial derivative) \emph{does} vanish may lead to additional physical solutions. The \verb|caseplot| routine for case tree plotting, with its ability display of zero and nonzero pivots for the cases, may then be used to visualize possible case trees that arise. 
      
  \item The choice of the constitutive dependencies \eqref{eq:def:constit} may have a significant effect on both the results themselves and the related symbolic computations. We note the importance of choosing a \emph{coordinate-invariant} approach. Indeed, if one allows the constitutive functions $\psi = \psi \left(\phi_C\right)$ \eqref{eq:def:constit} to depend on spatial partial derivatives, the dependence should in fact involve only coordinate-invariant combinations of such derivatives, such as a divergence, norm of a gradient, etc., but not, for example, Cartesian components of a gradient. Such choices would ensure the consistency of the application of the procedure in different coordinate systems. In particular, when computations are done in curvilinear coordinates, one needs to explicitly include into $\phi_C$ the spatial variables, to accommodate for the scaling (Lam\'{e}) factors that will appear in coordinate-invariant differential operators. The importance of this coordinate invariance requirement is illustrated in \cite{wang_shearing_1999};  in contrast, the original work of Liu \cite{liu_method_1972,liu2002shih} does not mention this aspect.

%

\end{enumerate}

It is also of interest to investigate the mathematical details related to the Liu's lemma itself, when it is applied to a set of nonlinear partial differential equations. The standard Liu algorithm described in Section \ref{sec:Mul-Liu} above attempts to apply Lemma \ref{lem:Liu} to move away from the solution set of the given PDE system \eqref{eq:gen:PhysPDEs}, and consider the vector $Y$ of higher-order partial derivatives of field variables as a set of arbitrary values. However, it is not completely clear how the linear algebra-based Lemma \ref{lem:Liu} applies to PDEs. Indeed, there is a substantial difference between linear equations and inequalities, where solutions and isosurfaces are given by linear or affine spaces, and solution manifolds of nonlinear differential equations in the jet spaces (see, e.g., \cite{olver2000applications}). It is normally assumed that the considered PDE systems are \emph{locally solvable}, i.e., the solution set of the PDE system in jet space is actually represented by these PDEs. In addition to the model PDEs \eqref{eq:gen:PhysPDEs}, further relationships between the field variables are provided by their \emph{differential consequences}. It follows that if one was to consistently follow through with the Liu's approach, the extended entropy inequality $\widetilde{\Pi}^{\eta}$ would possibly need to include the differential consequences of the equations $\Pi^{\phi}$ \eqref{eq:gen:PhysPDEs}, with additional multipliers. For example, for the heat-conducting fluid model \eqref{eq:fluid:gen}, if one extends the constitutive dependence $\phi_C$ \eqref{eq:simplefluid:constit:dependence} to include $\partial_t \rho$, it is clear that the time derivative $\partial_t \epsilon$ in the energy balance will include the second derivative $\partial_t^2 \rho$. It follows that the entropy condition \eqref{eq:Liu:Extended:entropy:ineq:RunEx} should include an additional term $\Lambda^{\rho}_1 \, \partial_t \Pi^{\rho}$ involving a time differential consequence of the continuity equation. In a similar way, spatial differential consequences $\partial_{x_i} \Pi^{\rho}$ would be added, with appropriate Lagrange multipliers.

In a related forthcoming publication, we will present a generalization of the Liu's algorithm, which considers the entropy inequality a priori on the solution space of dynamic equations of the model, thus requiring no Lagrange multipliers, and respecting the nonlinear nature of the solution set of the model.

\section*{Acknowledgement}

The authors thank Y. Wang and M. Oberlack for the ideas and discussions related to this project, and are also grateful for the financial support through the NSERC Discovery grant RGPIN-2014-05733, and the Deutsche Forschungsgemeinschaft (DFG) project WA 2610/3-1.

\bibliography{Bib_Entropy21c}

\bibliographystyle{ieeetr}

\end{document}